\pdfoutput=1
\documentclass[aps,prd,twocolumn,superscriptaddress,nofootinbib,floatfix]{revtex4-2}
\usepackage{amsmath,amssymb}
\usepackage{graphicx}
\usepackage{xcolor}
\usepackage[colorlinks=true,linkcolor=blue!60!black,citecolor=blue!60!black,urlcolor=blue!60!black]{hyperref}
\usepackage{bm}

\newcommand{\hMpc}{\,h\,\mathrm{Mpc}^{-1}}

\begin{document}
\raggedbottom

\title{\texorpdfstring{Partial geometrization of dark matter in metric $f(R)$ gravity:\\
Exact reconstruction and growth constraints}{Partial geometrization of dark matter in metric f(R) gravity: Exact reconstruction and growth constraints}}

\author{Davide Campanella Galanti}
\email{dcampanellagalanti@uninsubria.it}
\author{Oliver F. Piattella}
\email{of.piattella@uninsubria.it}
\affiliation{Dipartimento di Scienza e Alta Tecnologia (DiSAT), Universit\`a degli Studi dell'Insubria, via Valleggio 11, 22100 Como, Italy}
\affiliation{INFN, Sezione di Milano, via Celoria 16, 20133 Milano, Italy}

\date{\today}

\begin{abstract}
We ask what fraction of dark matter can be accounted for by the geometry of a stable metric $f(R)$ theory, requiring the expansion history to be exactly equal to that of the $\Lambda$CDM model. Only a fraction $\xi$ of dark matter clusters as dust, whereas $f(R)$ supplies the rest, as well as $\Lambda$. We show that the background reconstruction reduces to a linear equation for $F\equiv f_R$, whose solution is $F-1=(1-\xi)\,\Theta(a)$, with $\Theta$ independent of $\xi$: the departure from general relativity is exactly proportional to the geometrized fraction. We solve this equation in closed form in the matter$+\Lambda$ and radiation$+$matter eras: in both, the solution is an elementary particular solution plus two hypergeometric modes. Stability requires $F$ to decrease monotonically in time, so $F$ cannot take its general-relativity value $F=1$ both at early and at late times. The effective gravitational coupling $G/F$ therefore necessarily runs: between the radiation era and the de~Sitter attractor, $F$ changes by at least $0.837\,(1-\xi)$. In all stable reconstructions but one, $F-1$ grows as $R^{1/3}$ at high curvature, i.e., in dense environments such as the Sun, and local tests of gravity exclude them. The surviving member has $F\to1$ at early times: it recovers general relativity at high curvature and attains the minimal running of $F$. Keeping the linear growth factor at $k=0.1\hMpc$ within $10\%$ of its $\Lambda$CDM value requires $1-\xi\lesssim10^{-3}$, the weakest bound among all stable reconstructions; keeping $\sigma_8$ within $3\%$ requires $1-\xi\lesssim5\times10^{-5}$. Growth-rate data from DESI DR1 and SDSS DR16 give $1-\xi\lesssim10^{-3}$--$10^{-2}$, and the Cassini bound gives $1-\xi\lesssim10^{-10}$. A smoothly evolving metric $f(R)$ background can therefore geometrize only a negligible fraction of dark matter.
\end{abstract}

\maketitle

\section{Introduction}
\label{sec:intro}

Metric $f(R)$ gravity propagates an extra scalar degree of freedom, the scalaron, and the freedom in the choice of $f$ is enough to reproduce essentially any expansion history $H(a)$~\cite{Starobinsky1980,SotiriouFaraoni2010,DeFelice2010,Nojiri2006,Nojiri2017}. This reconstruction is not unique: for a given matter content, a whole class of $f(R)$ models shares the same expansion history~\cite{Multamaki2006}. For the $\Lambda$CDM background, the reconstruction has been worked out in closed form~\cite{delaCruzDombriz2006,Dunsby2010,He2013}. Reference~\cite{Dunsby2010} concluded that, with dust alone, the only real-valued $f(R)$ that reproduces this background is general relativity with a cosmological constant, $f=R-2\Lambda$, whereas Ref.~\cite{He2013} later found nontrivial real-valued solutions in terms of hypergeometric functions, contradicting this conclusion. The viability of such reconstructions has been examined~\cite{Fay2007}, and in the ``designer'' approach $f(R)$ is obtained by integrating the background equations numerically for a prescribed $H(a)$, leaving a single free parameter that sets the present range of the scalaron~\cite{Song2007,Pogosian2008}. The same freedom has motivated the idea that curvature may play the role of dark matter, both in the cosmological background~\cite{Capozziello2006} and on galactic scales, where $f(R)$ halo solutions have been proposed to reproduce flat rotation curves without particle dark matter~\cite{Capozziello2007,Bohmer2008,Shojai2014} (but see Ref.~\cite{Usman2016}).

Reproducing $H(a)$, however, constrains only the total dark stress-energy~\cite{Kunz2009}. How this is split between ``dark matter'' and ``geometry'' can be determined by investigating how small perturbations evolve, since structure formation requires a pressureless component that clusters. In metric $f(R)$, a smoothly evolving scalaron modifies the Poisson equation and generates a gravitational slip, but it does not provide an independent source of clustering.\footnote{In mimetic gravity, by contrast, a constrained conformal mode of the metric behaves as dust also in the perturbative regime~\cite{Chamseddine2013}.} The question we address is therefore quantitative. Suppose that a fraction $1-\xi$ of the dark matter density is traded for geometry, i.e., for an effective geometric density that scales as $a^{-3}$, like the cold dark matter it replaces, while $H(a)$ is kept exactly equal to its $\Lambda$CDM form. How large can $1-\xi$ be before the growth of structure departs from observations? The analogous question for a geometric (dark) radiation component was answered in Ref.~\cite{Morais2015}, and growth bounds on a matter-tracking curvature component in $R^{1+\epsilon}$ gravity were obtained in Ref.~\cite{Park2011}.

Our main results are the following. (i) As in Ref.~\cite{He2013}, the exact background reconstruction is linear in $F\equiv f_R$. Partial geometrization adds to the equation for $F-1$ a source proportional to the geometrized dust density, $(1-\xi)\rho_{\rm dm}$, so that $F-1$ is exactly proportional to $1-\xi$. For matter$+\Lambda$ and for radiation$+$matter, we give the solutions in closed form, so that $F$ and $f$ are explicit in each era (Sec.~\ref{sec:background}). (ii) Stability, $f_{RR}>0$, implies that $F$ decreases monotonically in time, so $F$ cannot equal 1 both at early and at late times: geometrization necessarily comes with a running effective gravitational coupling $G/F$. Once regularity at the de~Sitter attractor is imposed, the stable reconstructions form a one-parameter family, labeled by the present value $F_0$, whose members run in different ways. We study the whole family and, in particular, its two natural members: $F_0=1$ (N1) and $F\to1$ in the radiation era (N2). Only N2 recovers general relativity at high curvature, and local gravity excludes all the others (Sec.~\ref{sec:recloc}). (iii) For $F\geq1$, a scalaron enhancement held constant at its maximal value gives only the weak bound $\xi\geq0.70$ (Sec.~\ref{sec:optimistic}). On the exact backgrounds, the bounds are much stronger. Linear growth at $k\sim0.1\hMpc$ limits the geometrized fraction to $1-\xi\lesssim10^{-3}$ for every stable reconstruction. Requiring $\sigma_8$ within $10\%$ ($3\%$) of its $\Lambda$CDM value tightens this to $2\times10^{-4}$ ($5\times10^{-5}$), and a first comparison with DESI DR1 and SDSS DR16 growth-rate data gives $1-\xi\lesssim10^{-3}$--$10^{-2}$ (Secs.~\ref{sec:exactgrowth} and \ref{sec:data}). (iv) At these bounds, we expect the primary CMB anisotropies to change by less than $10^{-3}$. Local tests are far more restrictive: the Sun is screened only for $1-\xi\lesssim2\times10^{-6}$, and the Cassini bound requires $1-\xi\lesssim10^{-10}$ (Sec.~\ref{sec:recloc}).

Throughout we use a flat background with parameters close to the Planck 2018 values~\cite{Planck2018}: $\Omega_{\rm b}=0.050$, $\Omega_{\rm dm}=0.265$, $\Omega_{\rm m}=0.315$, $a_{\rm eq}=\Omega_{\rm r}/\Omega_{\rm m}=1/3400$, $\Omega_\Lambda=1-\Omega_{\rm m}-\Omega_{\rm r}$, $\sigma_8^{\Lambda{\rm CDM}}=0.811$, $n_{\rm s}=0.965$, and ${h=0.674}$ where needed. Recombination is at $a_{\rm rec}=1/1090$, matter-$\Lambda$ equality at $a_{{\rm m}\Lambda}=(\Omega_{\rm m}/\Omega_\Lambda)^{1/3}\simeq0.77$, and $\kappa=8\pi G$.

\section{Setup}
\label{sec:setup}

The action is
\begin{equation}
S=\frac{1}{2\kappa}\int d^4x\sqrt{-g}\,f(R)+S_{\rm m} ,
\label{eq:action}
\end{equation}
with field equations
\begin{equation}
FR_{\mu\nu}-\tfrac12 fg_{\mu\nu}-\nabla_\mu\nabla_\nu F+g_{\mu\nu}\Box F=\kappa T_{\mu\nu},
\label{eq:field}
\end{equation}
where $F\equiv f_R$ and $f_{RR}\equiv d^2f/dR^2$. On a flat Friedmann--Robertson--Walker background, $R={6(\dot H+2H^2)}$ and
\begin{align}
3FH^2&=\kappa\rho+\tfrac12(RF-f)-3H\dot F,\label{eq:fr1}\\
-2F\dot H&=\kappa(\rho+p)+\ddot F-H\dot F.\label{eq:fr2}
\end{align}
We impose the standard viability conditions on $f(R)$
\begin{equation}
F>0,\qquad f_{RR}>0,\qquad M^2\equiv\frac13\left(\frac{F}{f_{RR}}-R\right)>0,
\label{eq:stab}
\end{equation}
which keep the effective Newton constant positive, avoid the Dolgov--Kawasaki instability~\cite{DolgovKawasaki2003}, and keep the scalaron nontachyonic~\cite{DeFelice2010,Amendola2007}.

Our aim is to reproduce exactly the $\Lambda$CDM expansion history,
\begin{align}
\frac{H^2}{H_0^2}&\equiv E^2=\Omega_{\rm r}a^{-4}+\Omega_{\rm m}a^{-3}+\Omega_\Lambda,\notag\\
\frac{R}{H_0^2}&=3\Omega_{\rm m}a^{-3}+12\Omega_\Lambda,
\label{eq:target}
\end{align}
where radiation does not enter $R$ because its energy-momentum tensor is traceless, and $\Lambda\equiv3H_0^2\Omega_\Lambda$. The matter action $S_{\rm m}$ describes radiation and a clustering dust component, made of the baryons and of the fraction $\xi$ of dark matter that is not geometrized,
\begin{equation}
\rho_{\rm cl}=\rho_{\rm b}+\xi\rho_{\rm dm}=\epsilon_{\rm c}(\xi)\,\rho_{\rm m},\qquad \epsilon_{\rm c}\equiv\beta+\xi\gamma,
\label{eq:rhocl}
\end{equation}
where the label ``cl'' stands for clustering, $\beta=\Omega_{\rm b}/\Omega_{\rm m}$, and $\gamma=\Omega_{\rm dm}/\Omega_{\rm m}$, so that $\epsilon_{\rm c}$ is the fraction of all matter that clusters and $1-\epsilon_{\rm c}=\gamma(1-\xi)$. Both the missing fraction $(1-\xi)\rho_{\rm dm}$ and $\Lambda$ are supplied by $f(R)$: collecting the curvature terms of Eq.~\eqref{eq:fr1} with the bare coupling,
\begin{align}
3H^2&=\kappa(\rho_{\rm cl}+\rho_{\rm r})+\kappa\rho_{\rm eff},\notag\\
\kappa\rho_{\rm eff}&\equiv\tfrac12(RF-f)-3H\dot F+3H^2(1-F).
\label{eq:rhoeff}
\end{align}
Equation~\eqref{eq:target} requires $\rho_{\rm eff}=\rho_\Lambda+(1-\xi)\rho_{\rm dm}$ at all times, i.e., exactly $\Lambda$ plus dust: apart from $\rho_\Lambda$, the effective density scales as $a^{-3}$, like cold dark matter. The expansion history is therefore the same for every $\xi$: the parameter $\xi$ only determines how much of the matter clusters, and its effects appear in the perturbations.

\section{Exact background reconstruction}
\label{sec:background}

\subsection{\texorpdfstring{A linear equation for $F$}{A linear equation for F}}
\label{sec:lineq}

With $N=\ln a$ and $\dot F=HF_N$, Eq.~\eqref{eq:fr2} becomes, taking into account Eq.~\eqref{eq:rhocl},
\begin{equation}
F_{NN}+\left(\frac{H_N}{H}-1\right)F_N+2\frac{H_N}{H}F+3\epsilon_{\rm c}\Omega_{\rm m}(a)+4\Omega_{\rm r}(a)=0,
\label{eq:Feq}
\end{equation}
with $\Omega_{\rm m}(a)\equiv\Omega_{\rm m}a^{-3}/E^2$ and $\Omega_{\rm r}(a)\equiv\Omega_{\rm r}a^{-4}/E^2$. Given $F(a)$, the Friedmann equation~\eqref{eq:fr1}, in the form of Eq.~\eqref{eq:rhoeff}, determines $f$ algebraically,
\begin{equation}
f=RF-6H^2(F+F_N)+2\kappa(\rho_{\rm cl}+\rho_{\rm r}),
\label{eq:falg}
\end{equation}
and Eq.~\eqref{eq:Feq} guarantees that $df/dN=F\,dR/dN$. Since $R$ decreases monotonically along the $\Lambda$CDM history, $f(a)$ and $R(a)$ define a single-valued function $f(R)$, whose derivative is indeed $f_R=F$.

Equation~\eqref{eq:Feq} is linear. For $\xi=1$, without geometrization, it is solved by general relativity, $F=1$, since $2H_N/H=-3\Omega_{\rm m}(a)-4\Omega_{\rm r}(a)$. For $\xi<1$, $F-1$ obeys the same linear equation with the geometrized dust, $3\gamma(1-\xi)\Omega_{\rm m}(a)$, as its only source, and is therefore exactly proportional to $1-\xi$. We thus write $F=1+(1-\xi)\Theta$, where $\Theta(a)$ is the departure of $F$ from general relativity divided by the geometrized fraction $1-\xi$; it obeys
\begin{equation}
\Theta_{NN}+\left(\frac{H_N}{H}-1\right)\Theta_N+2\frac{H_N}{H}\Theta=3\gamma\,\Omega_{\rm m}(a),
\label{eq:Theta}
\end{equation}
which does not contain $\xi$. Equation~\eqref{eq:Theta} has the same form as the equation for $F-1$ in Ref.~\cite{He2013}, which is homogeneous for the exact $\Lambda$CDM background; partial geometrization adds the source on its right-hand side. If the boundary conditions do not depend on $\xi$, as for $F_0=1$ or $F\to1$ for $a\to0$ (Sec.~\ref{sec:stability}), $\Theta$ is a universal profile, i.e., the same for every $\xi$, and the whole family of backgrounds is described by
\begin{equation}
F(a;\xi)=1+(1-\xi)\,\Theta(a),\qquad f_{RR}=\frac{F_N}{R_N}=(1-\xi)\frac{\Theta_N}{R_N}.
\label{eq:universal}
\end{equation}
If the boundary conditions depended on $\xi$, each value of $\xi$ would select a different member of the family of Sec.~\ref{sec:stability}, and the analysis below would apply to each member separately. Inserting Eq.~\eqref{eq:universal} into Eq.~\eqref{eq:falg}, with $\rho_{\rm cl}=[1-\gamma(1-\xi)]\rho_{\rm m}$ and $6H^2=2\kappa(\rho_{\rm m}+\rho_{\rm r})+2\Lambda$, shows that the reconstructed function is also exactly linear in the geometrized fraction,
\begin{equation}
f=R-2\Lambda+(1-\xi)\left[R\,\Theta-6H^2(\Theta+\Theta_N)-2\gamma\kappa\rho_{\rm m}\right],
\label{eq:flin}
\end{equation}
where $f$ is given as a function of $R$ parametrically, through $a$ and $R(a)$ of Eq.~\eqref{eq:target}.

The solutions of Eq.~\eqref{eq:Theta} are simplest to describe era by era. Within each era $H_N/H$ is nearly constant, and the homogeneous solutions are power laws $a^s$ with $s^2+(H_N/H-1)s+2H_N/H=0$: $s=4,-1$ in the radiation era, $s=(5\pm\sqrt{73})/4\simeq3.39,-0.886$ in the matter era, and $s=0,1$ in the $\Lambda$ era. The source adds the particular solutions $-(\gamma/2)\,a/a_{\rm eq}$, $-\gamma$, and $[\gamma\Omega_{\rm m}/(4\Omega_\Lambda)]\,a^{-3}$, respectively.

\subsection{\texorpdfstring{Closed form for matter$+\Lambda$}{Closed form for matter + Lambda}}
\label{sec:closed}

At late times, $a\gg a_{\rm eq}$, radiation is negligible, and for $\Omega_{\rm r}\to0$ the reconstruction can be carried out in closed form. One has $H^2=R/3-\Lambda$ and $\dot R=-3H(R-4\Lambda)$, and Eq.~\eqref{eq:fr1} has the particular solution $f=\epsilon_{\rm c}(R-2\Lambda)$: general relativity with a rescaled Newton constant $G/\epsilon_{\rm c}$. By itself, this solution is observationally indistinguishable from $\Lambda$CDM: a constant $F=\epsilon_{\rm c}$ only rescales the Newton constant to $G/\epsilon_{\rm c}$, and the product $(G/\epsilon_{\rm c})\,\rho_{\rm cl}=G\rho_{\rm m}$ is the same as in $\Lambda$CDM, both in the background and in the perturbations. Geometrization becomes physical only through the homogeneous modes, which make $F$ time dependent, so that the value of $G$ measured locally differs from the one that governs cosmology. The remainder $u\equiv f-\epsilon_{\rm c}(R-2\Lambda)$ obeys, in the variable $x\equiv R/\Lambda-3=1+(\Omega_{\rm m}/\Omega_\Lambda)a^{-3}$ of Ref.~\cite{Dunsby2010}, the following equation:
\begin{equation}
x(1-x)\,u_{xx}+\frac{x-3}{6}\,u_x+\frac{u}{6}=0,
\label{eq:hyp}
\end{equation}
a Gauss hypergeometric equation with parameters $\alpha_\pm=(-7\pm\sqrt{73})/12$ and third parameter $-1/2$. As $a$ grows, $x$ decreases monotonically, from $x\gg1$ in the matter era to $x\to1$ at late times, the de~Sitter point $R=4\Lambda$ that the $\Lambda$CDM background approaches asymptotically; today $x_0=1+\Omega_{\rm m}/\Omega_\Lambda\simeq1.46$. The solutions about $x=\infty$~\cite{AbramowitzStegun}, already given in Eq.~(32) of Ref.~\cite{He2013}, give the general reconstructed $f(R)$
\begin{align}
f(R)&=\epsilon_{\rm c}(R-2\Lambda)+A\,w_+(x)+B\,w_-(x),\notag\\
w_\pm(x)&=x^{-\alpha_\pm}\,{}_2F_1\!\left(\alpha_\pm,\alpha_\pm+\tfrac32;1\pm\tfrac{\sqrt{73}}{6};x^{-1}\right).
\label{eq:wpm}
\end{align}
The homogeneous equation~\eqref{eq:hyp} is the one that governs the exact $\Lambda$CDM reconstruction with the full dust density~\cite{delaCruzDombriz2006,Dunsby2010,He2013}, while partial geometrization only adds the particular solution. In the matter era, $w_+\propto R^{-\alpha_+}$ gives $F-\epsilon_{\rm c}\propto a^{(5+\sqrt{73})/4}\simeq a^{3.39}$, while $w_-\propto R^{-\alpha_-}$ gives $F-\epsilon_{\rm c}\propto a^{(5-\sqrt{73})/4}\simeq a^{-0.886}$, a mode that grows into the past.

About the regular singular point $x=1$, the solutions of Eq.~\eqref{eq:hyp} behave as $(x-1)^0$ and $(x-1)^{2/3}$. The second one contributes to $F$ a term $\propto(x-1)^{-1/3}\propto a$, which diverges at the de~Sitter point and, unless its amplitude vanishes, drives either $f_{RR}$ or $F$ through zero in the future. Regularity at the de~Sitter attractor selects the one-parameter family
\begin{align}
f(R)&=\epsilon_{\rm c}(R-2\Lambda)+C\,{}_2F_1\!\left(\alpha_+,\alpha_-;\tfrac13;1-x\right),\notag\\
F&=\epsilon_{\rm c}+\frac{C}{2\Lambda}\,{}_2F_1\!\left(\alpha_++1,\alpha_-+1;\tfrac43;1-x\right).
\label{eq:dSreg}
\end{align}
The hypergeometric series in Eq.~\eqref{eq:dSreg} converges for $|1-x|<1$, i.e., for $x<2$, which corresponds to $a>a_{{\rm m}\Lambda}$, and at earlier times the functions are understood as their analytic continuation. Like $w_\pm$, this continuation is real for all $x\geq1$. Equation~\eqref{eq:dSreg} coincides, up to a normalization factor, with the first solution about $R=4\Lambda$ in Eq.~(35) of Ref.~\cite{He2013}. There, the authors retain the combination that vanishes at high curvature, for which $F$ and $f_{RR}$ diverge as $R\to4\Lambda$, but requiring regularity at the de~Sitter attractor excludes it. Continued analytically to $x\to\infty$~\cite{AbramowitzStegun}, the de~Sitter-regular solution~\eqref{eq:dSreg} becomes $A\,w_++B\,w_-$, with
\begin{equation}
\begin{gathered}
\frac{A}{C}=\frac{\Gamma(\frac13)\,\Gamma(\alpha_--\alpha_+)}{\Gamma(\alpha_-)\,\Gamma(\frac13-\alpha_+)}\simeq0.454,\\
\frac{B}{C}=\frac{\Gamma(\frac13)\,\Gamma(\alpha_+-\alpha_-)}{\Gamma(\alpha_+)\,\Gamma(\frac13-\alpha_-)}\simeq0.362,
\end{gathered}
\label{eq:AB}
\end{equation}
where the constant $C$ is fixed by the present value $F_0$. Therefore, requiring regularity at the de~Sitter attractor fixes $A/B\simeq1.254$, and the corresponding family of solutions contains the past-growing mode $a^{-0.886}$, unless $C=0$. In terms of $\varphi_0$ of Eq.~\eqref{eq:phi0} below, $C=2\Lambda(1-\xi)(\varphi_0+\gamma)/{}_2F_1(\alpha_++1,\alpha_-+1;\tfrac43;1-x_0)$, again proportional to $1-\xi$, and $F_0=1$ gives $C\simeq0.908\times2\Lambda\gamma(1-\xi)$. If radiation is taken into account, this closed form still describes the late-time profile of N1: it reproduces $F-\epsilon_{\rm c}$ to $0.1\%$ for $a\geq10^{-2}$. On N2, where $F-\epsilon_{\rm c}$ is small at late times, it misses the radiation tail $-8\gamma a_{\rm eq}/a$ of Sec.~\ref{sec:rm}, and reproduces $F-1$ to $0.5\%$ but $F-\epsilon_{\rm c}$ only to $20\%$ for $a\geq0.1$. Equations~\eqref{eq:wpm} and \eqref{eq:dSreg} agree with the numerical solution of Eq.~\eqref{eq:Feq} (with $\Omega_{\rm r}=0$) to $\lesssim10^{-10}$ (Appendix~\ref{app:numerics}).

\begin{figure*}[t]
\centering
\includegraphics[width=\textwidth]{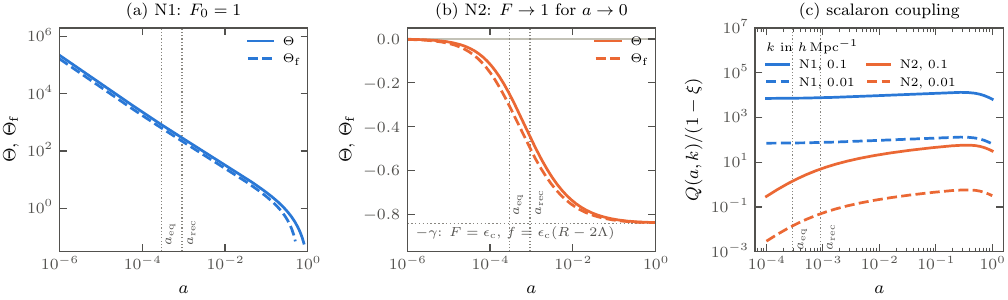}
\caption{Universal profiles of the de~Sitter-regular backgrounds: $\Theta=(F-1)/(1-\xi)$ (solid) and $\Theta_{\rm f}\equiv[f/(R-2\Lambda)-1]/(1-\xi)$ (dashed), i.e., the departures of $F$ and $f$ from their general-relativity values $F=1$ and $f=R-2\Lambda$, divided by the geometrized fraction $1-\xi$. (a) N1 ($F_0=1$): both grow as $a^{-1}$ in the radiation era, where $f-R\propto R^{4/3}$, while $\Theta_{\rm f}$ changes sign at $a\simeq0.6$ and is therefore shown only up to $a=0.5$. (b) N2 ($F\to1$ for $a\to0$): $F$ and $f/(R-2\Lambda)$ decrease from their general-relativity values to $\simeq\epsilon_{\rm c}=1-\gamma(1-\xi)$, i.e., to the particular solution $f=\epsilon_{\rm c}(R-2\Lambda)$, during the early matter era. The closed forms of Table~\ref{tab:explicit} coincide with these curves within the accuracies quoted there. (c) Scalaron coupling $Q=(k/a)^2f_{RR}/F$ divided by $1-\xi$, for $k=0.1$ (solid) and $0.01\hMpc$ (dashed). Vertical lines mark $a_{\rm eq}$ and $a_{\rm rec}$.}
\label{fig:background}
\end{figure*}

\subsection{\texorpdfstring{Closed form for radiation$+$matter}{Closed form for radiation + matter}}
\label{sec:rm}

Equation~\eqref{eq:dSreg} describes the matter-$\Lambda$ transition exactly, and the radiation-matter transition can be treated in the same way. For $a\ll a_{{\rm m}\Lambda}$ the cosmological constant is negligible in $H_N/H$ and in $\Omega_{\rm m}(a)$, and, with $y\equiv a/a_{\rm eq}$, Eq.~\eqref{eq:Theta} becomes
\begin{equation}
2y^2(1+y)\,\Theta_{yy}-y(4+3y)\,\Theta_y-2(4+3y)\,\Theta=6\gamma y .
\label{eq:rmeq}
\end{equation}
Like Eq.~\eqref{eq:hyp}, it has three regular singular points, here $y=0,-1,\infty$, and can therefore be reduced to the Gauss hypergeometric equation. The differential operator on the left-hand side of Eq.~\eqref{eq:rmeq} maps $1$ to $-2(4+3y)$ and $y^{-1}$ to $1$, so that a combination of the two reproduces the source $6\gamma y$, and a particular solution is elementary,
\begin{equation}
\Theta_{\rm p}=-\gamma\left(1+\frac{8}{y}\right).
\label{eq:rmpart}
\end{equation}
The homogeneous solution that vanishes as $a^4$ in the past is $y^4\,{}_2F_1(m_-,m_+;6;-y)$, with $m_\pm=(11\pm\sqrt{73})/4$, and the corresponding contribution to $f$, $\propto{}_2F_1(m_--1,m_+-1;5;-y)$, is the early-time solution of Ref.~\cite{He2013}. The solutions that reduce to the matter-era modes $a^{s_\pm}$ for $y\gg1$ are
\begin{equation}
\mathcal{T}_\pm(y)=y^{s_\pm}\,{}_2F_1\!\left(m_\mp,m_\mp-5;1\mp\tfrac{\sqrt{73}}{2};-y^{-1}\right).
\label{eq:rmhom}
\end{equation}
For $y<1$ they are understood as their analytic continuation. For $y\to0$ both grow as $a^{-1}$, $y\mathcal{T}_\pm\to K_\pm$, where, from the connection formula of ${}_2F_1$~\cite{AbramowitzStegun}, $K_\pm=24\,\Gamma(1\mp\tfrac{\sqrt{73}}{2})/[\Gamma(\tfrac{11\mp\sqrt{73}}{4})\,\Gamma(\tfrac{13\mp\sqrt{73}}{4})]$, with numerical values $K_+=8.4976$ and $K_-=0.99058$. In $\Theta=\Theta_{\rm p}+c_-\mathcal{T}_-+c_+\mathcal{T}_+$ the amplitude of the radiation-era mode, $\Theta\to K_1a_{\rm eq}/a$, is $K_1=c_-K_-+c_+K_+-8\gamma$. De~Sitter regularity makes $c_+$ of order $10^{-13}(\varphi_0+\gamma)$, negligible for $a\lesssim0.1$, and $c_-$ is fixed by the present value $F_0$ (Sec.~\ref{sec:stability}). Dropping $\mathcal{T}_+$, we have $F=1+(1-\xi)(\Theta_{\rm p}+c_-\mathcal{T}_-)$. Inserting it into Eq.~\eqref{eq:flin}, with $R=R_{\rm eq}y^{-3}$, where $R_{\rm eq}\equiv3H_0^2\Omega_{\rm m}a_{\rm eq}^{-3}$, and $6H^2=2R(1+y^{-1})$, gives the reconstruction for radiation$+$matter in closed form,
\begin{align}
f(R)&=R-2\Lambda-(1-\xi)\,R\left[\gamma\left(1+\frac{6}{y}\right)-c_-y^3\,\mathcal{U}(y)\right],\notag\\
F&=1-(1-\xi)\left[\gamma\left(1+\frac{8}{y}\right)-c_-\mathcal{T}_-(y)\right],
\label{eq:rmf}
\end{align}
where $y=(R_{\rm eq}/R)^{1/3}$ and
\begin{align}
\mathcal{U}(y)&=\frac{\sqrt{73}-7}{2}\,y^{s_--3}\notag\\
&\quad\times{}_2F_1\!\left(m_+-1,m_+-5;1+\tfrac{\sqrt{73}}{2};-y^{-1}\right)
\label{eq:rmU}
\end{align}
is the contribution of $\mathcal{T}_-$ to $f$, so that $f$ is an explicit function of $R$, with $df/dR=F$. In the matter era the $O(1-\xi)$ parts of both Eqs.~\eqref{eq:rmf} and \eqref{eq:dSreg} reduce to ${-\gamma(1-\xi)(R-2\Lambda)}$ plus the modes $a^{s_\pm}$, so that the two closed forms overlap for $a_{\rm eq}\ll a\ll a_{{\rm m}\Lambda}$.

With radiation, matter, and $\Lambda$ together, Eq.~\eqref{eq:Theta} has regular singular points at $a=0$, at $a=\infty$, and at the four zeros of $a^4E^2$, with Frobenius exponents $0$ and $1/2$. With six singular points it is no longer of hypergeometric or Heun type, and we know of no closed form. Since $a_{\rm eq}/a_{{\rm m}\Lambda}\simeq4\times10^{-4}$, however, the two transitions are well separated, and the two closed forms can be combined: their sum, minus their common matter-era part ($-\gamma$ plus the $a^{s_-}$ term of Eq.~\eqref{eq:dSreg}), reproduces $F-\epsilon_{\rm c}$ and $f_{RR}$ for $10^{-7}\leq a\leq1$ to $2\times10^{-4}$ on N1, and to $3\%$ and $5\%$, respectively, on N2, where $F-\epsilon_{\rm c}$ is itself small at late times. In what follows we integrate Eq.~\eqref{eq:Feq} numerically (Appendix~\ref{app:numerics}).

\subsection{\texorpdfstring{Stability, boundary conditions, and the running of $F$}{Stability, boundary conditions, and the running of F}}
\label{sec:stability}

Since $F_N=f_{RR}R_N$ and $R_N<0$ (Sec.~\ref{sec:lineq}),
\begin{equation}
f_{RR}>0\iff F_N<0.
\label{eq:monotonic}
\end{equation}
In a stable theory, $F$ therefore decreases monotonically in time. The second-order equation~\eqref{eq:Feq} requires two boundary conditions. The first is regularity at the de~Sitter attractor, i.e., analyticity of $f(R)$ at $R=4\Lambda$, which removes the future-growing mode ($F\propto a$ in the $\Lambda$ era), and for the values of $F_0$ allowed below we find that it also guarantees $f_{RR}>0$ at all times. The Compton parameter today, $B_0\equiv6H_0^2f_{RR}(a=1)/F_0$, which equals that of Ref.~\cite{Song2007} up to corrections of order $\Omega_{\rm r}/\Omega_{\rm m}$, is then an output rather than a free parameter, and is proportional to $1-\xi$ at first order. The second condition fixes the present value $F_0$, which we parametrize by
\begin{equation}
\varphi_0\equiv\frac{F_0-1}{1-\xi}=\Theta(a=1).
\label{eq:phi0}
\end{equation}
Every reconstruction with $\varphi_0\geq\varphi_0^{\rm N2}\simeq-0.837$ satisfies the conditions~\eqref{eq:stab} for every $0\leq\xi<1$ and at all times; we checked this for $10^{-8}\leq a\leq10^3$, although for very large $\varphi_0$ the margin of $M^2>0$ at the de~Sitter point becomes small. The lower end, $\varphi_0^{\rm N2}=\Theta_{\rm N2}(1)$, is the present value of the profile that recovers general relativity in the radiation era (N2 below). Like $\Theta$, it is fixed by the background and does not depend on $\xi$, and it is close to $-\gamma$ because $\Theta_{\rm N2}$ settles on the matter-era particular solution before $\Lambda$ takes over. For $\varphi_0<\varphi_0^{\rm N2}$ the radiation-era mode enters with negative amplitude: $f_{RR}$ becomes negative in the past and, further back, so does $F$. Two members of this family are singled out by natural boundary conditions:

\emph{N1: $F_0=1$ ($\varphi_0=0$).} The cosmological Newton constant today equals the bare $G$. The radiation-era mode $F\propto a^{-1}$, which grows into the past, is necessarily present: for $a\ll a_{\rm eq}$, $F-1\simeq756\,(1-\xi)\,a_{\rm eq}/a$, and correspondingly $f\simeq R-2\Lambda+567\,(1-\xi)\,R_{\rm eq}^{-1/3}R^{4/3}$. Today $B_0\simeq0.44\,(1-\xi)$. Since $R\propto a^{-3}$, $F-1\propto R^{1/3}$ keeps growing at high curvature, as it does for every member with $\varphi_0>\varphi_0^{\rm N2}$, and Sec.~\ref{sec:recloc} shows that local gravity excludes this behavior.

\emph{N2: $F\to1$ for $a\to0$ ($\varphi_0=\varphi_0^{\rm N2}$, the lower end of the family).} This boundary condition, usually adopted for designer models and in the early-time reconstruction of Ref.~\cite{He2013}, recovers general relativity in the radiation era and, more generally, at high curvature. The radiation-era past-growing mode is absent: for $a\ll a_{\rm eq}$, $F-1\simeq-\tfrac12\gamma(1-\xi)\,a/a_{\rm eq}$, and correspondingly $f\simeq R-2\Lambda-\tfrac34\gamma(1-\xi)\,R_{\rm eq}^{1/3}R^{2/3}$. Today $B_0\simeq2.1\times10^{-3}(1-\xi)$. Without radiation, requiring $F$ to stay bounded as $a\to0$ would set $C=0$ in Eq.~\eqref{eq:dSreg}, i.e., $f=\epsilon_{\rm c}(R-2\Lambda)$, so that on N2 all the nontrivial dynamics is driven by the radiation-matter transition.

\begin{figure}[b]
\centering
\includegraphics[width=\columnwidth]{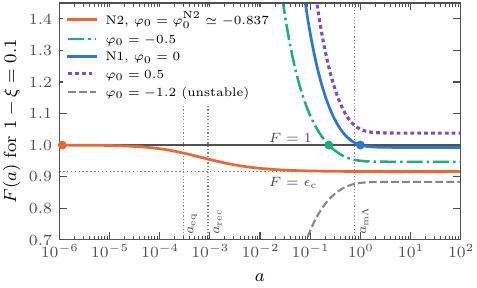}
\caption{Background $F(a)$ for $1-\xi=0.1$ along the family of de~Sitter-regular reconstructions, labeled by $\varphi_0$ of Eq.~\eqref{eq:phi0}. All members with $\varphi_0\geq\varphi_0^{\rm N2}$ are stable, while the dashed one, below the bound, is not. $F$ decreases monotonically and equals 1 at most once (dots): as $a\to0$ on N2, at $a\simeq0.24$ for $\varphi_0=-0.5$, and today on N1. For $\varphi_0<\varphi_0^{\rm N2}$, $F$ increases in time in the past (for the member shown, $f_{RR}<0$ at all times) and becomes negative further back. The dotted horizontal line is $F=\epsilon_{\rm c}$.}
\label{fig:family}
\end{figure}

\begin{figure}[t]
\centering
\includegraphics[width=\columnwidth]{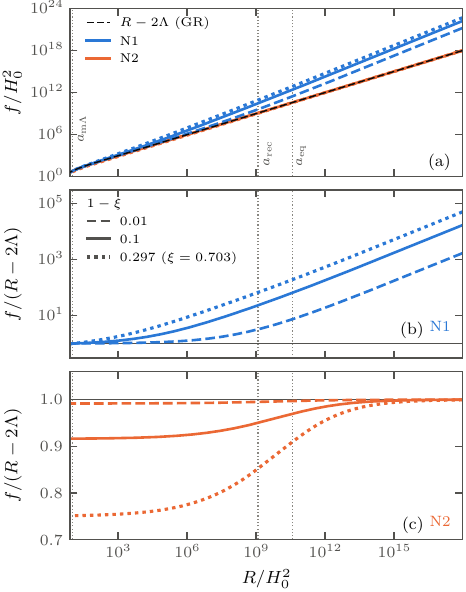}
\caption{Reconstructed $f(R)$ for $1-\xi=0.01$ (dashed), $0.1$ (solid), and $0.297$ (dotted), the largest geometrized fraction allowed by the optimistic bound, Eq.~\eqref{eq:optimistic}. (a) $f$ itself: on this scale N2 is indistinguishable from the general-relativity form $R-2\Lambda$ (black dashed). (b), (c) $f$ relative to $R-2\Lambda$ on N1 and N2. On N1, $f-R\propto R^{4/3}$ at high curvature, while on N2, $f\to R-2\Lambda$ at high curvature and $f\simeq\epsilon_{\rm c}(R-2\Lambda)$ at late times. The left edge is the de~Sitter point $R=4\Lambda$ (today $R\simeq1.11\times4\Lambda$), and vertical lines mark the curvature at $a_{\rm eq}$, $a_{\rm rec}$, and matter-$\Lambda$ equality $a_{{\rm m}\Lambda}$.}
\label{fig:fR}
\end{figure}

In terms of the modes of Sec.~\ref{sec:lineq}, de~Sitter regularity removes the $s=1$ mode and N2 removes the $s=-1$ mode, while on N1 the amplitude of the latter is fixed by $\Theta(1)=0$. In the closed form, N2 corresponds to $K_1=0$, i.e., $c_-=8\gamma/K_-\simeq6.794$. This value is fixed without reference to late times and reproduces $F-\epsilon_{\rm c}$ to $10^{-6}$ up to $a=10^{-2}$ and to $10^{-3}$ up to $a=0.1$. N2 is thus fixed by the early universe: $c_-$ does not depend on $\Lambda$, which affects the profile only at late times, through $H(a)$ and a negligible admixture of the $a^{3.39}$ mode. On N1, $\Theta(1)=0$ gives $K_1=755.51$ and $c_-=(8\gamma+K_1)/K_-=769.4874$. Matching the amplitudes of the modes $a^{s_\pm}$ of Eqs.~\eqref{eq:dSreg} and \eqref{eq:rmf} in the matter era gives $K_1\simeq753$, within $0.4\%$ of the numerical value, and the difference comes from the tail $-8\gamma a_{\rm eq}/a$ of Eq.~\eqref{eq:rmpart}, which decays almost as fast as $a^{s_-}$. Table~\ref{tab:explicit} collects the explicit $F$ and $f$ on N1 and N2.

The universal profiles of $F$ and $f$ are shown in Fig.~\ref{fig:background}. For N1, $\Theta_{\rm N1}(a_{\rm rec})\simeq286$ and $\Theta_{\rm N1}\propto a^{-1}$ in the radiation era. For N2, $\Theta_{\rm N2}$ decreases from $0$ to $\Theta_{\rm N2}(1)\simeq-0.837$, close to $-\gamma$, and $f$ moves accordingly from $R-2\Lambda$ to $\simeq\epsilon_{\rm c}(R-2\Lambda)$. Correspondingly, the effective coupling $G/F$ grows from $G$ to $\simeq G/\epsilon_{\rm c}$, which compensates for the missing clustering matter in the Friedmann equation. The change is $30\%$ complete at $a_{\rm eq}$, $53\%$ at $a_{\rm rec}$, $88\%$ at $a=0.01$, and $98\%$ at $a=0.1$, so it spans the whole early matter era.

Since $F$ decreases monotonically, its run between the radiation era and the de~Sitter attractor is smallest on N2, where $F$ goes from $1$ to $1+(1-\xi)\Theta_{\rm N2}(\infty)$, with $\Theta_{\rm N2}(\infty)\simeq\Theta_{\rm N2}(1)\simeq-0.837$, while every other member adds the radiation-era mode, which grows into the past. Across the family, therefore,
\begin{equation}
|\Delta F|\geq0.837\,(1-\xi).
\label{eq:DeltaF}
\end{equation}
Figure~\ref{fig:family} illustrates this for $1-\xi=0.1$. $F$ equals 1 at most once: asymptotically as $a\to0$ on N2, today on N1, and at an intermediate epoch for $\varphi_0^{\rm N2}<\varphi_0<0$ ($a\simeq0.24$ for ${\varphi_0=-0.5}$). No stable member has $F=1$ both early and late. Geometrizing dark matter in metric $f(R)$ therefore implies a time-dependent effective gravitational coupling, whose variation is of the same order as the geometrized fraction. This is the origin of all the constraints below.

\begin{table*}[t]
\caption{Explicit forms of $F$ and $f$ on the two exact backgrounds. At all times $F=1+(1-\xi)\Theta$ and Eq.~\eqref{eq:flin} hold exactly, and all figures use $\Theta$ from the numerical solution of Eq.~\eqref{eq:Theta}. The closed forms reproduce the numerical solution to the relative accuracy given in brackets, evaluated on the parts beyond the particular solution, $F-\epsilon_{\rm c}$ and $f-\epsilon_{\rm c}(R-2\Lambda)$, for $a\leq10^{-2}$ (radiation$+$matter) and for $10^{-2}\leq a\leq1$ on N1 and $0.1\leq a\leq1$ on N2 (matter$+\Lambda$). On N2 the matter$+\Lambda$ form misses the radiation tail $-8\gamma a_{\rm eq}/a$ of Eq.~\eqref{eq:rmpart}, which is comparable to $F-\epsilon_{\rm c}$ at late times, and the composite form of Sec.~\ref{sec:rm} reduces the error to $3\%$. Here $y=a/a_{\rm eq}=(R_{\rm eq}/R)^{1/3}$ and $x=R/\Lambda-3$; $\mathcal{T}_-$ and $\mathcal{U}$ are defined in Eqs.~\eqref{eq:rmhom} and \eqref{eq:rmU}; $\mathcal{G}(x)={}_2F_1(\alpha_++1,\alpha_-+1;\tfrac43;1-x)$, $\mathcal{J}(x)={}_2F_1(\alpha_+,\alpha_-;\tfrac13;1-x)$, and $b=(\varphi_0+\gamma)/\mathcal{G}(x_0)$, as in Eq.~\eqref{eq:dSreg}.}
\label{tab:explicit}
\begin{ruledtabular}
\begin{tabular}{llll}
Epoch & Closed form & N1 & N2\\
\hline
radiation, N1 & $F=1+756\,(1-\xi)\,y^{-1}$ & leading order & \\
 & $f=R-2\Lambda+567\,(1-\xi)R_{\rm eq}^{-1/3}R^{4/3}$ & & \\
radiation, N2 & $F=1-\tfrac12\gamma(1-\xi)\,y$ & & leading order\\
 & $f=R-2\Lambda-\tfrac34\gamma(1-\xi)R_{\rm eq}^{1/3}R^{2/3}$ & & \\
radiation$+$matter & $F=1-(1-\xi)\big[\gamma(1+8y^{-1})-c_-\mathcal{T}_-(y)\big]$ & $c_-=769.4874$ [$10^{-6}$] & $c_-=8\gamma/K_-$ [$10^{-6}$]\\
 & $f=R-2\Lambda-(1-\xi)R\big[\gamma(1+6y^{-1})-c_-y^3\,\mathcal{U}(y)\big]$ & [$4\times10^{-6}$] & [$4\times10^{-5}$]\\
matter$+\Lambda$ & $F=\epsilon_{\rm c}+(1-\xi)\,b\,\mathcal{G}(x)$ & $b=0.764$ [$10^{-3}$] & $b=3.95\times10^{-3}$ [$0.2$]\\
 & $f=\epsilon_{\rm c}(R-2\Lambda)+2\Lambda(1-\xi)\,b\,\mathcal{J}(x)$ & [$5\times10^{-4}$] & [$0.2$]\\
\end{tabular}
\end{ruledtabular}
\end{table*}

Figure~\ref{fig:fR} shows the reconstructed function itself for $1-\xi=0.01$, $0.1$, and $0.297$, the largest geometrized fraction allowed by the optimistic bound of Sec.~\ref{sec:optimistic}. On N2, $f$ approaches $R-2\Lambda$ at high curvature and $\epsilon_{\rm c}(R-2\Lambda)$ during the matter era. On N1, $f-R\propto R^{4/3}$ at high curvature, and at recombination $f$ exceeds $R-2\Lambda$ by factors of $3.2$, $23$, and $66$ for the three fractions, respectively. The geometrized fraction enters the growth of structure through $\epsilon_{\rm c}/F$: on large scales $Q\ll1$ and $\mu_\Psi=\Sigma=1/F$ (Sec.~\ref{sec:growth}), so $\epsilon_{\rm c}/F$ is the source of the Poisson equation relative to $\Lambda$CDM. On N2 it rises from $\epsilon_{\rm c}$ to unity across the radiation-matter transition. On N1 it vanishes in the past: at recombination it is $0.26$, $0.03$, and $0.009$ for the same three fractions.

\subsection{Effective geometric stress-energy}
\label{sec:Tgeom}

\begin{figure*}[t]
\centering
\includegraphics[width=\textwidth]{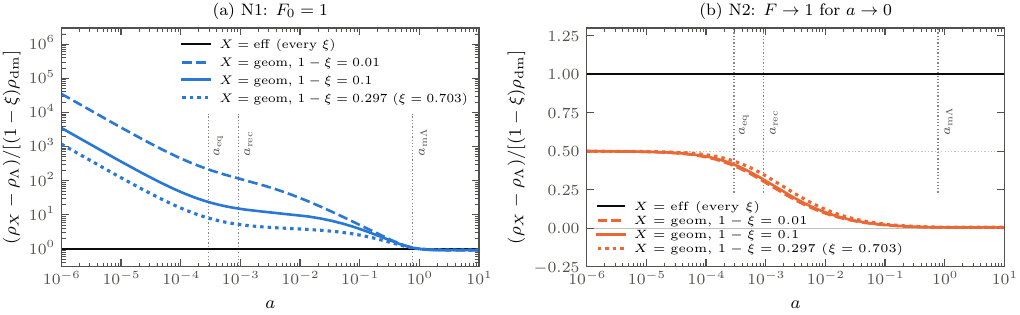}
\caption{Geometric densities relative to the geometrized dust, $(\rho_X-\rho_\Lambda)/[(1-\xi)\rho_{\rm dm}]$, for the two definitions of Sec.~\ref{sec:Tgeom}: the effective density of Eq.~\eqref{eq:rhoeff}, $X={\rm eff}$, for which the ratio equals 1 for every $\xi$ and $a$, and the density of Eq.~\eqref{eq:rhogeomratio}, $X={\rm geom}$, shown for $1-\xi=0.01$ (dashed), $0.1$ (solid), and $0.297$ (dotted). (a) N1; (b) N2. Vertical lines mark $a_{\rm eq}$, $a_{\rm rec}$, and $a_{{\rm m}\Lambda}$.}
\label{fig:densities}
\end{figure*}

The geometric density can be defined in two ways. The effective density $\rho_{\rm eff}$ of Eq.~\eqref{eq:rhoeff}, defined with the bare coupling, is separately conserved, because of the Bianchi identity and $\nabla^\mu T_{\mu\nu}=0$. Alternatively, dividing Eq.~\eqref{eq:field} by $F$ and writing $G_{\mu\nu}=\kappa(F^{-1}T_{\mu\nu}+T^{\rm geom}_{\mu\nu})$, one finds $\nabla^\mu T^{\rm geom}_{\mu\nu}=-T_{\mu\nu}\nabla^\mu(F^{-1})$ and $\kappa\rho_{\rm geom}=\Lambda+\kappa\rho_{\rm m}(1-\epsilon_{\rm c}/F)+\kappa\rho_{\rm r}(1-1/F)$, which is neither constant nor proportional to $a^{-3}$. The two definitions differ by $(\rho_{\rm cl}+\rho_{\rm r})(1-1/F)$. Since $F-1\propto1-\xi$, this difference is comparable to the geometrized density $(1-\xi)\rho_{\rm dm}$ or larger, except near $a=1$ on N1, where $F\simeq1$. Requiring $\rho_{\rm geom}$, rather than $\rho_{\rm eff}$, to scale as $a^{-3}$ would therefore not reproduce Eq.~\eqref{eq:target}. Using $1-\epsilon_{\rm c}/F=(1-\xi)(\Theta+\gamma)/F$ and $1-1/F=(1-\xi)\Theta/F$, the geometric density relative to the geometrized dust is
\begin{equation}
\frac{\rho_{\rm geom}-\rho_\Lambda}{(1-\xi)\rho_{\rm dm}}=\frac{\Theta+\gamma+(a_{\rm eq}/a)\,\Theta}{\gamma F},
\label{eq:rhogeomratio}
\end{equation}
while the same ratio is exactly 1 for $\rho_{\rm eff}$ (Fig.~\ref{fig:densities}). On N2 the geometric density is half of $(1-\xi)\rho_{\rm dm}$ in the radiation era, where $\Theta\simeq-(\gamma/2)\,a/a_{\rm eq}$, and nearly vanishes in the matter era, where $\Theta\simeq-\gamma$: there the missing dust is supplied not by a geometric density but by the larger coupling $G/F$. On N1 it exceeds $(1-\xi)\rho_{\rm dm}$ by orders of magnitude in the past and approaches it only today. The dust-like behavior of $\rho_{\rm eff}$ holds only for the background, while in the perturbations the geometric sector does not cluster like dust (Sec.~\ref{sec:growth}).

\section{Perturbations and growth}
\label{sec:growth}

\subsection{Quasi-static limit}

\begin{figure*}[t]
\centering
\includegraphics[width=\textwidth]{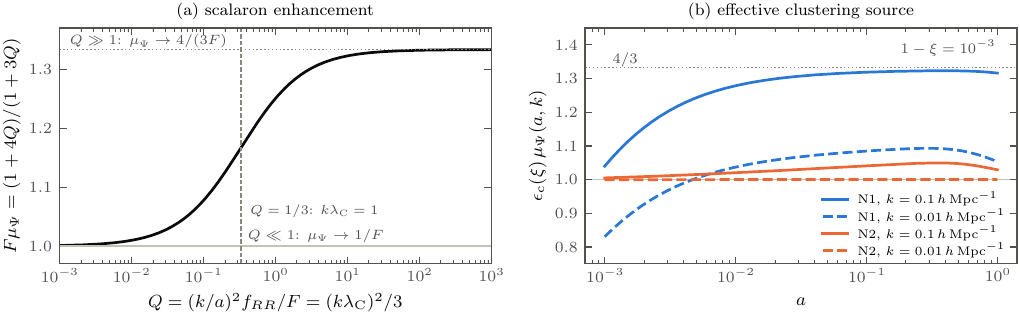}
\caption{(a) Scalaron enhancement of the Newtonian potential, $F\mu_\Psi=(1+4Q)/(1+3Q)$, Eq.~\eqref{eq:mu}, as a function of $Q=(k\lambda_{\rm C})^2/3$: it goes from 1 on scales larger than the Compton wavelength to $4/3$ well inside it. (b) Effective clustering source $\epsilon_{\rm c}\mu_\Psi$ relative to $\Lambda$CDM, for which it equals 1, for $1-\xi=10^{-3}$. On N1 the source is suppressed at early times on large scales ($F>1$) and approaches the ceiling $4/3$ at late times on small scales, while on N2 it exceeds unity through the scalaron term, increasingly so at small scales.}
\label{fig:mu}
\end{figure*}

We perturb the trace of Eq.~\eqref{eq:field}, $3\Box F+RF-2f=\kappa T$, in Newtonian gauge, $ds^2=-(1+2\Psi)dt^2+a^2(1-2\Phi)d\mathbf{x}^2$, with $\delta F=f_{RR}\delta R$. This gives a massive, driven wave equation for the scalaron,
\begin{multline}
\delta\ddot{F}+3H\delta\dot{F}+\left(\frac{k^2}{a^2}+M^2\right)\delta F=\frac{\kappa}{3}\,\delta\rho_{\rm cl}\\
+2\Psi\ddot F+\dot F\left(6H\Psi+3\dot\Phi+\dot\Psi\right),
\label{eq:dF}
\end{multline}
where $M^2$ is the mass of Eq.~\eqref{eq:stab}. Since $M^2>0$ on all stable reconstructions, the homogeneous solutions of Eq.~\eqref{eq:dF} do not grow, and $\delta F$ is driven only by the matter and metric perturbations on its right-hand side. The trace-free spatial equation gives the slip, $F(\Phi-\Psi)=\delta F$. In what follows we restrict to subhorizon scales and assume $F/f_{RR}\gg R$. For $a\geq10^{-3}$ the latter holds with $Rf_{RR}/F\lesssim80\,(1-\xi)$ on N1 and $\lesssim0.06\,(1-\xi)$ on N2. Under these conditions the scalaron follows the matter adiabatically,
\begin{equation}
\delta F=\frac{\kappa f_{RR}\,\rho_{\rm cl}\,\delta_{\rm cl}}{F(1+3Q)},\qquad Q\equiv\frac{k^2}{a^2}\frac{f_{RR}}{F}.
\label{eq:dFqs}
\end{equation}
The scalaron perturbation is small compared with the matter perturbation: using $f_{RR}/F=Qa^2/k^2$, $\kappa\rho_{\rm cl}=3H^2\Omega_{\rm cl}(a)$, with $\Omega_{\rm cl}(a)\equiv\epsilon_{\rm c}\Omega_{\rm m}(a)$, and $Q/(1+3Q)<1/3$, one finds
\begin{equation}
\frac{|\delta F|}{F\delta_{\rm cl}}=\frac{Q}{1+3Q}\,\frac{\kappa\rho_{\rm cl}}{F}\,\frac{a^2}{k^2}\leq\frac{\Omega_{\rm cl}(a)}{F}\left(\frac{aH}{k}\right)^2\ll1 .
\label{eq:dFsmall}
\end{equation}
The geometric density perturbation is not small, but it is entirely determined by the clustering matter: in the same limit $\kappa\delta\rho_{\rm geom}=-(k/a)^2\delta F/F$, i.e., $\delta\rho_{\rm geom}=-[Q/(1+3Q)]\,\delta\rho_{\rm cl}/F$, which is anticorrelated with $\delta\rho_{\rm cl}$ and reaches $-\delta\rho_{\rm cl}/(3F)$ for $Q\gg1$. Its effective sound speed squared, $\delta p_{\rm geom}/\delta\rho_{\rm geom}\to-2/3$ on small scales, is not that of a physical fluid (cf.~Ref.~\cite{Arjona2019}): in this regime the geometric sector has no independent dynamics and does not cluster like the dust it replaces. The modified Poisson and lensing equations read~\cite{DeFelice2010,Tsujikawa2007}
\begin{equation}
\mu_\Psi(a,k)=\frac1F\,\frac{1+4Q}{1+3Q},\qquad \Sigma(a)=\frac1F.
\label{eq:mu}
\end{equation}
The scalaron multiplies the Newtonian potential of the clustering matter by $(1+4Q)/(1+3Q)$ [Fig.~\ref{fig:mu}(a)]: this factor tends to 1 on scales larger than the comoving Compton wavelength $\lambda_{\rm C}=a^{-1}\sqrt{3f_{RR}/F}$, where $Q=(k\lambda_{\rm C})^2/3\ll1$, and reaches $4/3$ well inside it. The lensing potential receives no scale-dependent (scalaron) enhancement. Keeping $r\equiv Rf_{RR}/F$ in $M^2$, i.e., without assuming $F/f_{RR}\gg R$, gives $\mu_\Psi=(1+4Q-r)/[F(1+3Q-r)]$, which changes the growth ratio $R_{\rm D}$ of Sec.~\ref{sec:exactgrowth} by less than $10^{-5}$ at our bounds and by at most $1\%$ for any $1-\xi\leq1$, even though $r$ reaches $0.3$ on N1 at large $1-\xi$. The linear growth factor $D$ of the clustering dust obeys
\begin{equation}
D_{NN}+\left(2+\frac{H_N}{H}\right)D_N-\frac32\,\epsilon_{\rm c}\,\Omega_{\rm m}(a)\,\mu_\Psi(a,k)\,D=0,
\label{eq:growth}
\end{equation}
where, as before, the subscript $N$ denotes $d/dN$. The combination $\epsilon_{\rm c}\mu_\Psi$ is the effective clustering source relative to $\Lambda$CDM [Fig.~\ref{fig:mu}(b)]. By Eq.~\eqref{eq:universal}, $Q\propto(1-\xi)k^2$ up to corrections of order $(1-\xi)^2$. The comoving Compton wavelength is therefore $\simeq(0.8$--$1.3)\times10^2\sqrt{1-\xi}\,h^{-1}\,\mathrm{Mpc}$ on N2 and $\simeq(1.4$--$2.0)\times10^3\sqrt{1-\xi}\,h^{-1}\,\mathrm{Mpc}$ on N1, for $0.01\leq a\leq1$, and on N2 it decreases to $40\sqrt{1-\xi}\,h^{-1}\,\mathrm{Mpc}$ at $a=10^{-3}$. Unlike standard designer models, whose scalaron coupling decays rapidly into the past, here the scalaron is active over the whole matter era [Fig.~\ref{fig:background}(c)], and for this reason the present-day parameters $B_0$ or $F_0-1$ alone cannot summarize these models.

\subsection{\texorpdfstring{An optimistic bound for $F\geq1$}{An optimistic bound for F >= 1}}
\label{sec:optimistic}

\begin{figure*}[t]
\centering
\includegraphics[width=\textwidth]{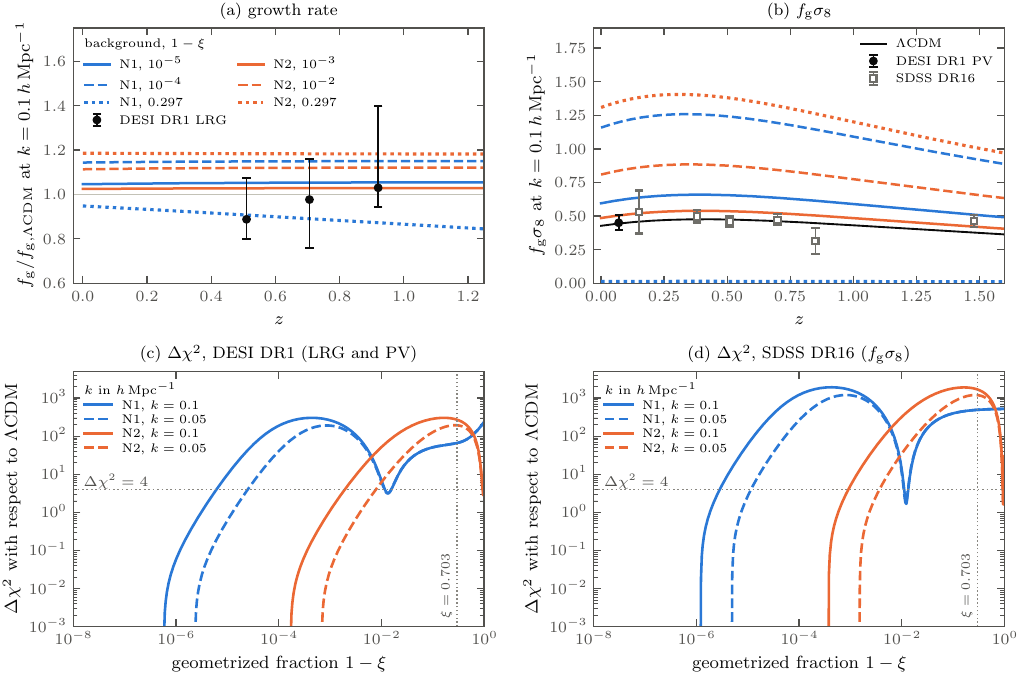}
\caption{(a) Growth rate relative to $\Lambda$CDM at $k=0.1\hMpc$ for representative geometrized fractions, including the optimistic bound $1-\xi=0.297$ (dotted), compared with the DESI DR1 LRG measurements~\cite{NovellMasot2025}. (b) $f_{\rm g}\sigma_8$ for the same cases, with the DESI DR1 peculiar-velocity point~\cite{Qin2026} and the SDSS DR16 consensus values~\cite{Alam2021}. (c), (d) $\Delta\chi^2$ with respect to $\Lambda$CDM as a function of $1-\xi$ for the DESI data (LRG and PV) and for the SDSS data, with $f_{\rm g}$ evaluated at $k=0.1$ (solid) and $0.05\hMpc$ (dashed). The dotted lines mark $\Delta\chi^2=4$ and the optimistic bound $\xi=0.703$ (derived for $F\geq1$), and the dips below $\Delta\chi^2=4$ near $1-\xi\simeq10^{-2}$ (N1) and $0.95$ (N2) correspond to the cancellation windows of Sec.~\ref{sec:exactgrowth}. At small $1-\xi$, $\Delta\chi^2$ is slightly negative (down to $-0.01$ for DESI and $-0.3$ for SDSS) and falls below the plotted range.}
\label{fig:desi}
\end{figure*}

If $F\geq1$, as on N1 until today, since $F$ decreases monotonically to $F_0=1$ [Eq.~\eqref{eq:monotonic}], then $\mu_\Psi\leq4/3$ for every $a$ and $k$. In matter domination and for constant $\epsilon_{\rm c}\mu_\Psi$, $D\propto a^n$ with $n=(-1+\sqrt{1+24\epsilon_{\rm c}\mu_\Psi})/4$, and the standard growing mode $n=1$ requires $\epsilon_{\rm c}\mu_\Psi=1$. With the most favorable value $\mu_\Psi=4/3$, this requires
\begin{equation}
\xi\geq\frac{3/4-\beta}{\gamma}\simeq0.703.
\label{eq:optimistic}
\end{equation}
The same condition, $g\equiv\epsilon_{\rm c}\mu_\Psi=1$, selects the standard growing mode in the other eras. For constant $g$, the growing mode is $D_+=a^n\,{}_2F_1\big(\tfrac n3,\tfrac{n+2}{3};\tfrac{2n}{3}+\tfrac76;-\tfrac{\Omega_\Lambda}{\Omega_{\rm m}}a^3\big)$ for matter$+\Lambda$ and $D_+=P_\nu\big(\sqrt{1+a/a_{\rm eq}}\big)$ for radiation$+$matter, where $P_\nu$ is the Legendre function of the first kind and $\nu=(-1+\sqrt{1+24g})/2$. Both reduce to the $\Lambda$CDM and M\'esz\'aros~\cite{Meszaros1974} modes for $g=1$. Equation~\eqref{eq:optimistic} thus holds in all eras. It is only an optimistic bound: it assumes a permanently unscreened scalaron and ignores the running of $F$. For the members with $F\geq1$, i.e., $\varphi_0\geq0$, it is a necessary condition for recovering the $\Lambda$CDM growth in the matter era. It depends, however, on the choice of $F_0$: when $F<1$, as on N2 at all times, it does not apply, because $\epsilon_{\rm c}/F\simeq1$ already restores the $\Lambda$CDM source in the matter era, and the constraint comes instead from the scale dependence of $\mu_\Psi$. On the exact backgrounds the value $1-\xi=0.297$ that saturates Eq.~\eqref{eq:optimistic} is excluded by a wide margin: at $k=0.1\hMpc$ it gives $R_{\rm D}\simeq0.03$ on N1 and $R_{\rm D}\simeq2.6$ on N2, far from the $\Lambda$CDM value $R_{\rm D}=1$ favored by the growth data of Sec.~\ref{sec:data} [see Fig.~\ref{fig:growth}(b) and the dotted curves of Fig.~\ref{fig:desi}].

\subsection{Growth on the exact backgrounds}
\label{sec:exactgrowth}

\begin{figure*}[t]
\centering
\includegraphics[width=\textwidth]{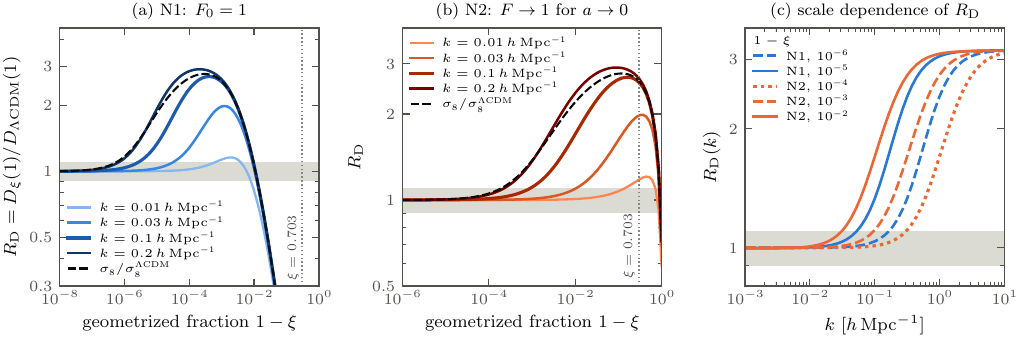}
\caption{(a), (b) Growth ratio $R_{\rm D}$, Eq.~\eqref{eq:RD}, as a function of the geometrized fraction $1-\xi$ for four scales, and $\sigma_8/\sigma_8^{\Lambda{\rm CDM}}$ (black dashed), on N1 (a) and N2 (b), with different horizontal ranges. The dotted vertical lines mark the optimistic bound $\xi=0.703$ of Eq.~\eqref{eq:optimistic}, derived for $F\geq1$. (c) $R_{\rm D}$ as a function of $k$ for representative values of $1-\xi$ on N1 (blue) and N2 (orange): it grows with $k$, since $Q\propto(1-\xi)k^2$, and saturates for $Q\gg1$. In all panels the gray band marks $|R_{\rm D}-1|\leq10\%$, i.e., a linear growth factor within $10\%$ of its $\Lambda$CDM value, the criterion of Table~\ref{tab:bounds}.}
\label{fig:growth}
\end{figure*}

On the backgrounds of Eq.~\eqref{eq:universal}, we integrate Eq.~\eqref{eq:growth} from $a_{\rm i}=10^{-3}$, with the $\Lambda$CDM (M\'esz\'aros) growing mode as initial condition, and define the growth ratio
\begin{equation}
R_{\rm D}(\xi,k)=\frac{D_\xi(a=1,k)}{D_{\Lambda{\rm CDM}}(a=1)}.
\label{eq:RD}
\end{equation}
This is the ratio of the growth factors for a common amplitude at $a_{\rm i}$, while differences accumulated before $a_{\rm i}$ belong to the recombination physics of Sec.~\ref{sec:recloc}. Since $\mu_\Psi$ depends on scale, so does $R_{\rm D}$. The amplitude $\sigma_8$ of matter fluctuations averages $R_{\rm D}$ over the linear power spectrum,
\begin{equation}
\frac{\sigma_8^2}{(\sigma_8^{\Lambda{\rm CDM}})^2}=\frac{\int d\ln k\,k^3P_{\Lambda{\rm CDM}}(k)\,W^2(kR_8)\,R_{\rm D}^2(k)}{\int d\ln k\,k^3P_{\Lambda{\rm CDM}}(k)\,W^2(kR_8)},
\label{eq:sigma8}
\end{equation}
with $R_8=8\,h^{-1}\,\mathrm{Mpc}$, the top-hat window $W(kR_8)=3[\sin(kR_8)-kR_8\cos(kR_8)]/(kR_8)^3$, and $P_{\Lambda{\rm CDM}}\propto k^{n_{\rm s}}T_{\rm nw}^2(k)$, where $T_{\rm nw}$ is the no-wiggle transfer function of Ref.~\cite{EisensteinHu1998}, and the weight peaks at $k\simeq0.2\hMpc$. Figure~\ref{fig:growth} shows $R_{\rm D}$ and $\sigma_8$ as functions of $1-\xi$ [panels (a) and (b)], and $R_{\rm D}$ as a function of $k$ [panel (c)].

On N1 two effects compete. The first is the early suppression $\mu_\Psi\simeq1/F$. The second is an enhancement close to its maximum, $\mu_\Psi\to4/3$ for $Q\gg1$, at $a\gtrsim0.01$, because the past-growing mode makes $f_{RR}$ large at late times [$Q/(1-\xi)\sim10^4$ at $k=0.1\hMpc$, Fig.~\ref{fig:background}(c)]. The enhancement dominates for $1-\xi\lesssim10^{-2}$ and the suppression at larger values. The two effects cancel in a narrow window near $1-\xi\simeq10^{-2}$, where $|R_{\rm D}-1|\leq10\%$ again, and its position depends on scale: $1-\xi\in[0.0039,0.0080]$ for $k=0.01\hMpc$, $[0.0089,0.0121]$ for $k=0.03\hMpc$, and $[0.0095,0.0127]$ for $k\geq0.1\hMpc$. Within the window $F(a_{\rm rec})\simeq2$--$5$, i.e., gravity at recombination is weaker by that factor (Sec.~\ref{sec:recloc}), and no value of $\xi$ passes the criterion at all scales. N1 is, in any case, excluded by local gravity. On N2 the scalaron enhancement dominates and grows as $(1-\xi)k^2$ until $Q\gg1$ on all linear scales. $R_{\rm D}$ then peaks, at $1-\xi\simeq0.16$ and $R_{\rm D}\simeq2.7$ for $k=0.1\hMpc$, and decreases again because the early suppression $\epsilon_{\rm c}/F<1$ (Sec.~\ref{sec:stability}) keeps growing with $1-\xi$. It returns to unity near $1-\xi\simeq0.91$--$0.93$ for $k\geq0.03\hMpc$ but near $0.80$ for $k=0.01\hMpc$, so that no value passes the criterion at all scales, and at these large fractions $\epsilon_{\rm c}/F(a_{\rm rec})\simeq0.4$.

\begin{table}[b]
\caption{Largest geometrized fraction $1-\xi$ below which $|R_{\rm D}-1|\leq10\%$ or $3\%$ at scale $k$, or, in the row $\sigma_8$, $|\sigma_8/\sigma_8^{\Lambda{\rm CDM}}-1|\leq10\%$ or $3\%$, on N1 and N2, while the narrow cancellation windows at larger $1-\xi$ are discussed in Sec.~\ref{sec:exactgrowth}. The last rows give the smallest $1-\xi$ with $\Delta\chi^2>4$ with respect to $\Lambda$CDM for the DESI DR1 and SDSS DR16 data of Sec.~\ref{sec:data}, with $f_{\rm g}$ evaluated at the quoted $k$.}
\label{tab:bounds}
\begin{ruledtabular}
\begin{tabular}{ccccc}
$k\,[h\,\mathrm{Mpc}^{-1}]$ & N1, $10\%$ & N1, $3\%$ & N2, $10\%$ & N2, $3\%$\\
\hline
$0.01$ & $6.3\times10^{-4}$ & $1.5\times10^{-4}$ & $0.14$ & $0.039$\\
$0.03$ & $3.2\times10^{-5}$ & $9.2\times10^{-6}$ & $1.1\times10^{-2}$ & $3.0\times10^{-3}$\\
$0.1$ & $2.7\times10^{-6}$ & $7.9\times10^{-7}$ & $9.2\times10^{-4}$ & $2.6\times10^{-4}$\\
$0.2$ & $6.8\times10^{-7}$ & $2.0\times10^{-7}$ & $2.3\times10^{-4}$ & $6.6\times10^{-5}$\\
$\sigma_8$ & $5.6\times10^{-7}$ & $1.6\times10^{-7}$ & $2.0\times10^{-4}$ & $5.4\times10^{-5}$\\
\hline
 & \multicolumn{2}{c}{N1, DESI} & \multicolumn{2}{c}{N2, DESI}\\
$0.05$ & \multicolumn{2}{c}{$2.5\times10^{-5}$} & \multicolumn{2}{c}{$7.8\times10^{-3}$}\\
$0.1$ & \multicolumn{2}{c}{$6.2\times10^{-6}$} & \multicolumn{2}{c}{$1.9\times10^{-3}$}\\
\hline
 & \multicolumn{2}{c}{N1, SDSS} & \multicolumn{2}{c}{N2, SDSS}\\
$0.05$ & \multicolumn{2}{c}{$1.2\times10^{-5}$} & \multicolumn{2}{c}{$3.7\times10^{-3}$}\\
$0.1$ & \multicolumn{2}{c}{$2.9\times10^{-6}$} & \multicolumn{2}{c}{$9.2\times10^{-4}$}\\
\end{tabular}
\end{ruledtabular}
\end{table}

The largest geometrized fraction below which $|R_{\rm D}-1|\leq10\%$ (or $3\%$) holds is given in Table~\ref{tab:bounds}. At $k=0.1\hMpc$,
\begin{equation}
1-\xi\lesssim3\times10^{-6}\ \ ({\rm N1}),\qquad 1-\xi\lesssim9\times10^{-4}\ \ ({\rm N2}).
\label{eq:bounds}
\end{equation}
The quasi-static approximation is least reliable on the largest scales at early times, since $k=0.01\hMpc$ crosses the horizon close to $a=10^{-3}$. Starting the integration at $a_{\rm i}=10^{-2}$ instead changes the bounds by less than $50\%$: those at $k\geq0.03\hMpc$ become weaker, and that at $k=0.01\hMpc$ tighter. For $k\gtrsim0.03\hMpc$ they scale approximately as $k^{-2}$, as expected from $Q\propto(1-\xi)k^2$, so the bound is set by the smallest linear scales [Fig.~\ref{fig:growth}(c)]. The amplitude $\sigma_8$, dominated by $k\simeq0.1$--$0.3\hMpc$, gives $1-\xi\lesssim2\times10^{-4}$ on N2 at $10\%$, and $\lesssim5\times10^{-5}$ at $3\%$. The latter criterion is comparable to the current precision on $\sigma_8$ and $S_8\equiv\sigma_8\sqrt{\Omega_{\rm m}/0.3}$, which is $1$--$2\%$ from CMB lensing~\cite{Madhavacheril2024,Qu2026} and $2$--$3\%$ from cosmic shear~\cite{Wright2025}. Geometrization enhances growth at these bounds, so it cannot ease a low-$S_8$ tension. Massive neutrinos suppress growth on the same scales and are partly degenerate with an $f(R)$ enhancement~\cite{Baldi2014}, but for the neutrino masses allowed by current data the suppression of $\sigma_8$ is at the percent level, so it can relax the $3\%$ numbers by at most a factor of order unity.

The adiabatic response of the scalaron, Eq.~\eqref{eq:dFqs}, can be tested directly. At the bounds of Table~\ref{tab:bounds}, $M/H\gtrsim4$ at the two N1 bounds for $k=0.01\hMpc$ and $M/H\gtrsim9$ at all the others, for $a\geq10^{-3}$. We have solved Eq.~\eqref{eq:dF} without the metric-potential terms, from quasi-static initial data at $a_{\rm i}$, and used the resulting $\delta F$ in the Poisson equation. This changes $R_{\rm D}$ by less than $10^{-3}$ at all the bounds of Table~\ref{tab:bounds}, and by less than $10^{-4}$ except for N1 at $k=0.01\hMpc$. The only remaining approximation is the subhorizon one, which matters only for the largest scale near $a_{\rm i}$, and its effect is bounded by the comparison with $a_{\rm i}=10^{-2}$ above.

We have repeated the analysis across the whole stable family, Eq.~\eqref{eq:phi0}. The bound tightens monotonically as $\varphi_0$ moves away from its minimum, so N2 is the most favorable stable reconstruction. The decrease is steep: at $k=0.1\hMpc$ the bound is $9.2\times10^{-4}$ on N2, $1.0\times10^{-4}$ for $\varphi_0=\varphi_0^{\rm N2}+0.02$, $7\times10^{-6}$ for $\varphi_0=-0.5$ (the member of Fig.~\ref{fig:family} that has $F=1$ at $a\simeq0.24$), $2.7\times10^{-6}$ on N1, and $1\times10^{-6}$ for $\varphi_0=1.5$. At larger $1-\xi$, the members of the family also show narrow cancellation windows analogous to the one on N1. The windows for $k\geq0.03\hMpc$ can overlap, but none of them overlaps the window for $k=0.01\hMpc$ (we checked twelve members with $\varphi_0^{\rm N2}\leq\varphi_0\leq5$), and all of them have $\epsilon_{\rm c}/F(a_{\rm rec})\lesssim0.4$: the dust density that gravitates at recombination would be reduced by more than $60\%$ (Sec.~\ref{sec:recloc}). With the criterion imposed at all linear scales, the N2 value in Eq.~\eqref{eq:bounds} therefore holds for every stable, de~Sitter-regular reconstruction of the $\Lambda$CDM background. Relaxing de~Sitter regularity on N2, i.e., admitting the mode $F\propto a$ of the $\Lambda$ era with an amplitude small enough to keep $f_{RR}>0$ until today, changes this bound by less than $3\%$, although such a mode drives $f_{RR}$ or $F$ through zero in the future.

\section{Comparison with growth-rate data}
\label{sec:data}

As a first test against data, we compare the growth rate $f_{\rm g}\equiv d\ln D/d\ln a$ with two sets of DESI DR1 measurements. The first set consists of the luminous red galaxy (LRG) values from the joint power spectrum and bispectrum analysis, $f_{\rm g}/f_{\rm g}^{\rm fid}=0.888^{+0.186}_{-0.089}$, $0.977^{+0.182}_{-0.220}$, $1.030^{+0.368}_{-0.085}$ at $z=0.510$, $0.706$, $0.920$~\cite{NovellMasot2025}, where $f_{\rm g}^{\rm fid}$ is the growth rate of the Planck 2018 $\Lambda$CDM cosmology, which we identify with our reference. The second is the peculiar-velocity (PV) consensus value $f_{\rm g}\sigma_8(z=0.07)=0.450\pm0.055$~\cite{Qin2026}, for which our $\Lambda$CDM reference gives $0.444$. As a cross-check, we use instead the SDSS DR16 consensus values of $f_{\rm g}\sigma_8$ from the joint analyses of baryon acoustic oscillations and redshift-space distortions, at six redshifts between $0.15$ and $1.48$~\cite{Alam2021}, which partly overlap in volume with the DESI data and are therefore not combined with them. These measurements probe the growth of structure on scales of tens of megaparsecs, the cosmological regime in which geometrization would have to replace the clustering dark matter. We evaluate $f_{\rm g}$ at $k=0.05$ and $0.1\hMpc$, typical of the scales that dominate redshift-space distortion fits, and compute $\sigma_8(z)=\sigma_8^{\Lambda{\rm CDM}}D_\xi(z)/D_{\Lambda{\rm CDM}}(z=0)$, with $D_\xi$ evaluated at the same $k$ and normalized to the same amplitude at $a_{\rm i}$. For each data set we compute $\chi^2=\sum_i(m_i-d_i)^2/\sigma_i^2$ for the measurements $d_i\pm\sigma_i$ and the model predictions $m_i$. For the asymmetric DESI errors we use the side on which the prediction lies, i.e., a split-normal approximation of the posteriors, and we neglect correlations between the measurements. The model is compared with $\Lambda$CDM through $\Delta\chi^2\equiv\chi^2(\xi)-\chi^2_{\Lambda{\rm CDM}}$ [Fig.~\ref{fig:desi}(c),(d)], and with one parameter $\Delta\chi^2>4$ excludes a value of $1-\xi$ at about $95\%$ confidence. This criterion is approximate, since $\Lambda$CDM lies at the boundary $1-\xi=0$ and the SDSS data are fit marginally better at small $1-\xi$ ($\Delta\chi^2_{\rm min}\simeq-0.3$), and a precise analysis would use the published likelihoods and covariances. $\Lambda$CDM itself has $\chi^2=0.5$ for the four DESI measurements and $6.2$ for the six SDSS ones.

On N2 the DESI data exclude $1-\xi>1.9\times10^{-3}$ ($\Delta\chi^2>4$) for $k=0.1\hMpc$ and $1-\xi>7.8\times10^{-3}$ for $k=0.05\hMpc$, in line with the growth criterion, while the SDSS data give $9.2\times10^{-4}$ and $3.7\times10^{-3}$. On N1 the DESI data exclude $1-\xi>6.2\times10^{-6}$ ($k=0.1\hMpc$) and $>2.5\times10^{-5}$ ($k=0.05\hMpc$). On both backgrounds the cancellations of Sec.~\ref{sec:exactgrowth} leave a narrow interval with $\Delta\chi^2<4$, $1-\xi\simeq0.011$--$0.016$ on N1 and $1-\xi\simeq0.93$--$0.99$ on N2, where $\Delta\chi^2_{\rm min}\simeq3.2$ and $2.9$ for the DESI data and $1.7$ and $1.6$ for the SDSS data, on N1 and N2, respectively. The data alone do not exclude these intervals, but the scale dependence of the growth, recombination, and local gravity do (Secs.~\ref{sec:exactgrowth} and \ref{sec:recloc}). At the optimistic bound, $\xi=0.703$, the DESI data at $k=0.1\hMpc$ give $\Delta\chi^2\simeq66$ on N1 and $270$ on N2. This comparison is deliberately simple: the scale-dependent growth of $f(R)$ is not captured by template fits with a single growth rate, and only four (DESI) or six (SDSS) measurements are used. A likelihood analysis including the DESI DR1 full-shape results~\cite{DESI2024V,Ishak2024} and the scale-dependent power spectrum would make these bounds quantitative.

\section{Recombination and local gravity}
\label{sec:recloc}

Since $H(a)$ is exactly that of $\Lambda$CDM, all background quantities are unchanged by construction: the expansion rate during nucleosynthesis, the sound horizon, and the distance to last scattering. Bounds on the gravitational constant that act through $H(T)$ therefore do not apply directly. The running of $F$ shows up instead in the perturbations and in local physics, and we estimate both effects here without a full treatment.

\emph{Recombination.} All species gravitate with $G/F$. On N1, $F(a_{\rm rec})=1+286\,(1-\xi)$: for $1-\xi=10^{-2}$ gravity at recombination is weaker by a factor $3.9$, which would remove most of the dark-matter-driven potential wells that shape the acoustic peaks. Keeping this change below one percent requires $1-\xi\lesssim3.5\times10^{-5}$. On N2, the effective gravitating dust density at recombination differs from its $\Lambda$CDM value by $1-\epsilon_{\rm c}/F(a_{\rm rec})\simeq0.40\,(1-\xi)$, and the photon-baryon fluid gravitates with $G/F(a_{\rm rec})\simeq G[1+0.44(1-\xi)]$. Compared with the percent-level precision of Planck on $\omega_{\rm c}\equiv\Omega_{\rm dm}h^2$~\cite{Planck2018}, this suggests $1-\xi\lesssim$ a few percent. At the bounds of Eq.~\eqref{eq:bounds}, $|F(a_{\rm rec})-1|\lesssim10^{-3}$ on both backgrounds, so we expect the primary anisotropies to change by less than $10^{-3}$, a much weaker constraint than the one from growth. CMB lensing, which follows the late-time growth, is covered by the $\sigma_8$ criterion, and the late integrated Sachs--Wolfe effect changes as little as the growth on large scales. Gravitational waves propagate at the speed of light in $f(R)$~\cite{Ezquiaga2017}, and the running of the effective Planck mass, $\alpha_{\rm M}=d\ln F/d\ln a=(1-\xi)\Theta_N/F$~\cite{BelliniSawicki2014}, is $|\alpha_{\rm M}|\lesssim10^{-6}$ today at our bounds. A Boltzmann treatment is required to make these estimates quantitative.

\emph{Local gravity.} In the chameleon picture~\cite{KhouryWeltman2004,HuSawicki2007,Faulkner2007}, an object of Newtonian potential $U_{\rm N}$ is screened only if the scalaron field differs between its interior and its surroundings by $\lesssim\frac23U_{\rm N}$. The reconstruction fixes $f(R)$ at all the curvatures reached along the $\Lambda$CDM history, and in a medium of density $\rho$ the field takes the cosmological value at the epoch at which the matter density was $\rho$. The Galactic environment, $\rho\sim10^6\rho_{\rm crit}$, corresponds to $a\simeq7\times10^{-3}$, and the mean density of the Sun to $a\simeq1.2\times10^{-10}$, i.e., $T\sim1$~MeV. In the background of Eq.~\eqref{eq:target}, where radiation is traceless, this curvature is still reached along the expansion history that fixes the reconstruction. On N2, $F\to1$ at high curvature, and the field differs by $0.13\,(1-\xi)$ between the cosmological background and the Milky Way, and by $0.71\,(1-\xi)$ between the Galactic environment and the interior of the Sun. With $U_{\rm N}\sim10^{-6}$ for the Milky Way and $U_{\rm N}\simeq2\times10^{-6}$ for the Sun, the thin-shell criterion requires $1-\xi\lesssim5\times10^{-6}$ and $\lesssim2\times10^{-6}$, respectively. Screening is necessary but not sufficient: outside a screened Sun the residual field shifts the post-Newtonian parameter by $|\gamma_{\rm PPN}-1|\simeq|\Delta F_\odot|/U_{\rm N}$, where $\Delta F_\odot$ is the field difference between the Galactic environment and the interior of the Sun, so that the Cassini bound, $|\gamma_{\rm PPN}-1|<2.3\times10^{-5}$~\cite{Bertotti2003}, requires $|\Delta F_\odot|\lesssim5\times10^{-11}$ for any $f(R)$~\cite{HuSawicki2007}. Since the scalaron range in the Galactic environment, $\simeq0.5\sqrt{1-\xi}\,h^{-1}\,\mathrm{Mpc}$, far exceeds the size of the solar system, the residual field is not Yukawa suppressed there, and the Cassini bound gives $1-\xi\lesssim7\times10^{-11}$ on N2. Every other member of the family contains the radiation-era mode, so that $F-1\simeq900\,(\varphi_0-\varphi_0^{\rm N2})(1-\xi)(R/R_{\rm eq})^{1/3}$ grows without bound at high curvature. On N1, $F-1\simeq47\,(1-\xi)$ in the Galactic environment and $\simeq1.8\times10^9(1-\xi)$ at the mean solar density: screening the Sun requires $1-\xi\lesssim10^{-15}$ and the Cassini bound $1-\xi\lesssim3\times10^{-20}$, and for a generic member of the family $(1-\xi)(\varphi_0-\varphi_0^{\rm N2})\lesssim2\times10^{-20}$. The Newton constant measured in the laboratory would also differ from the cosmological one by a factor $F(\rho_{\rm lab})\simeq4\times10^3$ at the N1 growth bound. Local gravity therefore selects N2, the reconstruction that recovers general relativity at high curvature. N2 also avoids the curvature singularity found in other $f(R)$ models that recover general relativity at high curvature~\cite{Frolov2008,KobayashiMaeda2008}: since $1-F\propto R^{-1/3}$, its scalaron potential diverges as $F\to1$, so the high-curvature limit cannot be reached at finite energy. These are order-of-magnitude estimates, which assume that the chameleon picture applies to the reconstructed $f(R)$ in static configurations. A dedicated analysis would also connect our results to the galactic $f(R)$ literature~\cite{Capozziello2007,Bohmer2008,Shojai2014,Usman2016}: the same $f(R)$ must pass both the cosmological and the local tests.

\section{Discussion and conclusions}
\label{sec:conclusions}

We have quantified what fraction of dark matter can be accounted for by the geometry of a stable metric $f(R)$ theory, in the form of an effective density that scales as $a^{-3}$ like cold dark matter, while the expansion history is kept exactly equal to that of the $\Lambda$CDM model. The background problem is linear. Its general solution for matter$+\Lambda$ is $f=\epsilon_{\rm c}(R-2\Lambda)$ plus the hypergeometric modes of the exact $\Lambda$CDM reconstruction, and $F-1$ is exactly proportional to the geometrized fraction. The particular solution shows what geometrization means in metric $f(R)$: a rescaling of the Newton constant by $1/\epsilon_{\rm c}$, which by itself is observationally indistinguishable from $\Lambda$CDM. Stability forces $F$ to decrease monotonically, so that it cannot equal unity both early and late, and $F$ runs by at least $0.837\,(1-\xi)$ between the radiation era and the de~Sitter attractor.

This running is what makes the scenario observable. For $F\geq1$, the argument based only on $\mu_\Psi\leq4/3$ gives $\xi\geq0.70$. On the exact backgrounds the bound is far stronger, and local gravity selects the boundary condition on $F$. Unless general relativity is recovered at high curvature, $F-1$ grows as $R^{1/3}$, and screening the Sun requires $(1-\xi)(\varphi_0-\varphi_0^{\rm N2})\lesssim10^{-15}$. For $F=1$ today (N1), where early gravity is suppressed and a light scalaron boosts late-time growth, linear growth at $k=0.1\hMpc$ alone would already require $1-\xi\lesssim3\times10^{-6}$. If general relativity holds at high curvature, the effective coupling rises during the early matter era and the scalaron-enhanced growth requires $1-\xi\lesssim10^{-3}$ at $k=0.1\hMpc$, and $\lesssim5\times10^{-5}$ for $\sigma_8$ within $3\%$. This reconstruction, N2, is the most favorable member of the whole stable family. DESI DR1 and SDSS DR16 growth-rate data give bounds of the same order, and the Cassini bound requires $1-\xi\lesssim7\times10^{-11}$. The clustering, dust-like component must therefore constitute essentially all of the dark matter. This parallels the conclusion reached for a geometric dark-radiation component in Ref.~\cite{Morais2015}, and it is the $f(R)$ counterpart of a lesson familiar from unified dark-matter models: a component that behaves as dust in the background can replace cold dark matter only if it also clusters like dust~\cite{Sandvik2004,Piattella2010,Bertacca2011}. In mimetic gravity this holds by construction: the constrained conformal mode of the metric behaves as pressureless dust also in the perturbations~\cite{Chamseddine2013}. In metric $f(R)$ the scalaron is instead massive and follows the matter adiabatically, Eq.~\eqref{eq:dFqs}, and cannot play this role.

Our conclusions apply to stable, de~Sitter-regular metric $f(R)$ theories that reproduce the $\Lambda$CDM expansion exactly. The main limitations are the following. We use the quasi-static approximation~\cite{delaCruzDombriz2008,SawickiBellini2015}. The scalaron follows it closely at all our bounds (Sec.~\ref{sec:exactgrowth}), while the subhorizon approximation is marginal only for $k=0.01\hMpc$ near $a_{\rm i}=10^{-3}$, and our bounds change by less than $50\%$ when starting at $a_{\rm i}=10^{-2}$. We restrict to linear scales, where the scalaron responds linearly to the matter, Eq.~\eqref{eq:dFqs}, and the chameleon mechanism, which requires large field excursions inside dense objects, does not operate, and we also restrict to smooth reconstructions, in which $F$ evolves on Hubble time scales. A scalaron oscillating about its minimum, proposed as a dark-matter candidate~\cite{Cembranos2009,Katsuragawa2017,Parbin2021}, clusters like dust and would belong to the fraction $\xi$. Nonlinear probes are beyond our scope. At the N2 growth bound, $F_0-1\simeq-8\times10^{-4}$ and $B_0\simeq2\times10^{-6}$. The field difference between the background and cluster interiors, $\simeq(0.01$--$0.04)(1-\xi)\approx10^{-5}$, is then comparable to cluster potentials, so clusters would be at most partially screened, and cluster abundances~\cite{Cataneo2015} may tighten the bound further. We do not perform a CMB analysis, but at our bounds we expect the effects on the primary anisotropies to be below the $10^{-3}$ level (Sec.~\ref{sec:recloc}). The natural continuation is a Boltzmann implementation. The geometric sector maps onto a designer $f(R)$ model whose effective dark energy tracks matter at early times, $\rho^{\rm eff}_{\rm DE}=\rho_\Lambda+(1-\xi)\rho_{\rm dm}$, and can be treated within effective-field-theory codes~\cite{Hu2014,Zumalacarregui2017}. This would turn the recombination estimates of Sec.~\ref{sec:recloc} into a quantitative CMB constraint, and it would allow a direct comparison with existing bounds on designer and chameleon models~\cite{Lombriser2012,Cataneo2015,Battye2018,Vogt2025}, which cannot be translated into bounds on $\xi$ through $B_0$ or $F_0-1$ alone.

\begin{acknowledgments}
Numerical results were obtained with NumPy, SciPy, mpmath, SymPy, and Matplotlib. The Python code that reproduces all figures and tables is available from the authors upon request.
\end{acknowledgments}
\appendix
\section{Numerical implementation}
\label{app:numerics}

Equation~\eqref{eq:Theta} is a linear two-point boundary-value problem. We integrate the de~Sitter-regular particular solution and the de~Sitter-regular homogeneous solution backward from $a=10^3$, starting from their asymptotic forms, $\Theta_\Lambda\simeq[\gamma\Omega_{\rm m}/(4\Omega_\Lambda)]\,a^{-3}$ and $\Theta_{\rm h}\simeq1+[\Omega_{\rm m}/(4\Omega_\Lambda)]\,a^{-3}$, with an eighth-order Runge--Kutta scheme (relative tolerance $10^{-13}$). The N2 solution is obtained by canceling the coefficient of the radiation-era mode $a^{-1}$, extracted at $a=10^{-10}$; any other member of the family follows as $\Theta=\Theta_{\rm N2}+(\varphi_0-\varphi_0^{\rm N2})\,\Theta_{\rm h}/\Theta_{\rm h}(1)$. We checked that (i) a collocation boundary-value solver with asymptotic boundary conditions at $a=10^{-8}$ and $a=10^3$ reproduces N2 to $10^{-13}$ at $a=1$; (ii) without radiation, the de~Sitter-regular homogeneous solution agrees with Eq.~\eqref{eq:dSreg} to $\lesssim10^{-10}$, and $F=\epsilon_{\rm c}$ is the particular solution; (iii) $f$ from Eq.~\eqref{eq:falg} satisfies $df/dN=FR_N$ to the accuracy of the numerical derivative; (iv) the conditions~\eqref{eq:stab} hold for $10^{-8}\leq a\leq10^3$ for all tested $\varphi_0\geq\varphi_0^{\rm N2}$ and $\xi\in[0,1)$, and all members reduce to $F\equiv1$ for $\xi=1$; (v) solving Eq.~\eqref{eq:Feq} directly for $F$ with $\xi=0.5$, $0.9$, $0.99$, and $0.999$ gives $(F-1)/(1-\xi)=\Theta_{\rm N1}$ to $5\times10^{-10}$; (vi) the connection coefficients $A/C$ and $B/C$ of Eq.~\eqref{eq:AB} agree with a direct evaluation of the connection formula for $2.5\leq x\leq10^3$, the radiation-matter solutions of Sec.~\ref{sec:rm} satisfy Eq.~\eqref{eq:rmeq} to 40 digits, and the closed forms of Table~\ref{tab:explicit} and their composite reproduce the numerical profiles (on N2, also the collocation solution) within the quoted accuracies; (vii) the perturbation of $\Box F$ entering Eq.~\eqref{eq:dF} agrees with a direct symbolic expansion; and (viii) at the bounds of Table~\ref{tab:bounds}, Eq.~\eqref{eq:dF} without the metric-potential terms, coupled to the growth equation, reproduces the quasi-static $\delta F$ to $\lesssim2\%$ for $a\geq10^{-3}$ and $\lesssim0.3\%$ for $a\geq10^{-2}$, and $R_{\rm D}$ to better than $10^{-3}$.
The growth equation~\eqref{eq:growth} is integrated with the same scheme (relative tolerance $10^{-10}$) from the M\'esz\'aros growing mode at $a_{\rm i}$, and we checked the closed-form growing modes of Sec.~\ref{sec:optimistic} against their defining equations. For $\sigma_8$ we compute $R_{\rm D}$ at 41 logarithmically spaced scales between $10^{-3}$ and $10\hMpc$ and interpolate with a cubic spline.


\end{document}